\documentclass{iaus}
\usepackage{graphicx}

\title[Properties of segregated star clusters] 
{\Large Integrated properties of mass segregated star clusters}

\author[Gaburov \& Gieles]   
{E. Gaburov$^{1,2}$\thanks{e-mail: egaburov@science.uva.nl, mgieles@eso.org} and 
  M. Gieles$^{3}$}

\affiliation{\footnotesize
$^1$Sterrenkundig Instituut ``Anton Pannekoek'', University of  Amsterdam \\
$^2$Section Computational Science, University of  Amsterdam \\
$^3$European Southern Observatory, Casilla 19001, Santiago 19, Chile 
}

\pubyear{2007}
\volume{246}  
\pagerange{1--2}
\date{?? and in revised form ??}
\jname{Proceedings Title IAU Symposium}
\editors{E. Vesperini (Chief Editor), M. Giersz, A. Sills}
\begin{document}

\maketitle

\begin{abstract}
In this contribution we study integrated properties of dynamically
segregated star clusters. The observed core radii of segregated
clusters can be 50\% smaller than the ``true'' core radius. In
addition, the measured radius in the red filters is smaller than those
measured in blue filters. However, these difference are small
($\lesssim10\%$), making it observationally challenging to detect mass
segregation in extra-galactic clusters based on such a comparison. Our
results follow naturally from the fact that in nearly all filters most
of the light comes from the most massive stars. Therefore, the
observed surface brightness profile is dominated by stars of similar
mass, which are centrally concentrated and have a similar spatial
distribution.

\end{abstract}

Mass segregation in star clusters is often observed from radial
variations in stellar mass function (\cite{dGr02a}; \cite{dGr02b};
\cite{kim06}). These methods are subject to biases and selection
effects, such as incompleteness and blending and can, therefore, not be
applied to clusters more distant than the Magellanic Clouds. Such
extra-galactic star clusters can only be studied through their
integrated properties (\cite{bastian07}; \cite{mccrady05}). In this
contribution we present simulated observations of integrated
properties of dynamically segregated young star clusters.


We present a semi-analytical model of star clusters based on a simple analytical description of the mass function at different radii from the cluster centre ($r$).
Based on both observations (\cite{kim06})
and $N$-body simulations (\cite{spz07arches}; \cite{eg07}) of young
star clusters, we model the mass function in the following way. For $r
< r_{\rm hm}$, with $r_{\rm hm}$ the half-mass radius, the mass function is
\[ 
  g(m,r < r_{\rm hm}) \propto  \left\{
  \begin{array}{ll}
    m^{\alpha_0}, & \textrm{if } m < \mu = 2\langle m\rangle\\
    \mu^{\alpha_0}\left({m\over \mu}\right)^{\alpha(r)}, & \textrm{otherwise}.
  \end{array}
  \right. ,
\]
whereas for $r > r_{\rm hm}$ it has power-law form with a constant
slope $\alpha_{\rm hm}$, $g(m, r > r_{\rm hm}) \propto m^{\alpha_{\rm
    hm}}$. Here, $\langle m\rangle$ and $\alpha_0$ are the mean mass
and the slope of the initial mass function respectively, and
$\alpha(r)$ is the index at the high-mass end which is a function of distance to
the cluster centre. We choose $\alpha(r) =
\alpha_{\inf} + (\alpha_c - \alpha_{\inf})(1 +
(r/\epsilon)^{\delta})$. The parameter $\alpha_c$ determines the
degree of mass segregation, while parameters $\epsilon$ and $\delta$
define shape of $\alpha(r)$. The two other parameters, $\alpha_{\rm
  hm}$ and $\alpha_{\inf}$, are determined by constraints that the
cluster integrated mass function results in IMF, and that the mean
mass is continuous at $r_{\rm hm}$.

Given the mass function, we can calculate the mean stellar mass,
$\mu(r)$, as well as the luminosity profile in different filters,
$L_\lambda(r)$, as a function of $r$. To this end we use
Padova isochrones (\cite{padova}). For a given  density
profile, $\rho(r)$, we can obtain spatial luminosity profile in a
desired filter by using the following conversion $L_\lambda(r) =
\rho(r)\cdot [\lambda_\lambda(r) / \mu(r)]$. As a test case, we choose
$\rho(r)$ to be an EFF profile (\cite{eff87}). The resulted
surface brightness profiles are fitted to a projected EFF profile
 in order to obtain the core radius and power-law index.

\begin{figure}
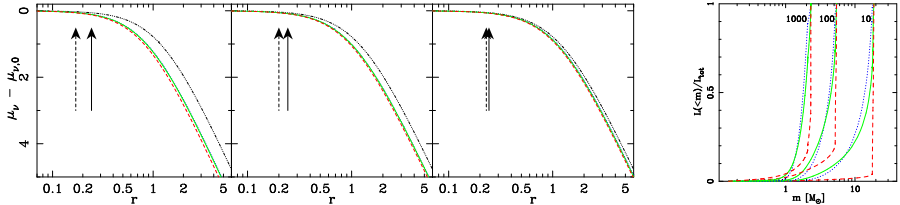

  \begin{center}
    \includegraphics[scale=0.45]{GaburovGieles_fig1.ps}$\qquad$
    \includegraphics[scale=0.155]{GaburovGieles_fig2.ps}
  \end{center}
  \caption{Normalised surface brightness profiles (left panel) of a
    star cluster as a function of its age. The black dash-dot-dot-dot
    line is the surface brightness profile of a non-segregated
    cluster, whereas the solid green, dotted blue and dashed red lines
    display the profiles for mass segregated clusters in the $V$, $U$
    and $K$ filters, respectively. The dashed and solid arrows show
    the core radius of segregated ($U$-filter) and non-segregated
    clusters respectively.  The cumulative luminosity functions (right
    panel) are computed from a single stellar population with a
    power-law mass function.  The numbers left of the lines show the
    age of the population in Myr.}
  \label{figure}
\end{figure}

In the case of non-segregated clusters, light distribution traces the
density distribution, which is not the case for segregated star
clusters. In the left panel of Fig.\,\ref{figure}, we show the surface
brightness profiles in different filters compared to surface
brightness profile of non-segregated cluster. We notice that the
observed core radius in segregated clusters is smaller by roughly 50\%
compared to the core radius of non-segregated cluster with the same
density profile. Moreover, the former one is never larger than the
latter one for ages $\lesssim$1\,Gyr. In addition, the difference
between core radii in $U$- and $K$-filters is smaller than 10\%, and
it is larger for younger clusters. 

Assuming a power-law mass function, we show the cumulative luminosity
function in different filters in the right panel of
Fig.\,\ref{figure}. In older clusters most of the light comes from
stars with similar masses which have light-to-mass ratio close to
unity. Therefore, the light distribution approximates well the density
distribution, and the core radii in different filters are nearly the
same. In the case of young star clusters, however, the most massive
stars dominate the $K$-filter, whereas stars with half the turn-off
mass dominate the $U$-filter. Combined with the fact that
light-to-mass ratio of these stars is notable large than unity, we
expected that core radius is smaller than that of unsegregated star
cluster. In addition, the core radius in $K$-filter is smaller than
the core radius in $U$-filter. However, we do not expect this
difference to be large, as the stars which differ in mass by a factor
of two do not have significantly different spatial distribution.

{\bf Acknowledgements:} We thank Nate Bastian and S{\o}ren Larsen for
helpful discussions. This work is supported by NWO under grant
\#635.000.303.

\end{document}